\documentclass[reprint,onecolumn,notitlepage,amsmath,amssymb,aps]{revtex4-1}

\usepackage{graphicx}
\usepackage{dcolumn}
\usepackage{bm}

\newcommand{\dprod}{\displaystyle\prod}
\newcommand{\tprod}{\textstyle\prod}

\begin{document}

\title{Renormalization-Group Stability Does Not Determine Front Selection}

\author{Ko Okumura}
 \altaffiliation{Department of Physics and Soft Matter Center, Ochanomizu University, 2-1-1, Ohtsuka, Bunkyo-ku, Tokyo 112-8610, Japan}

\date{\today}

\begin{abstract}

The Fisher-Kolmogorov-Petrovsky-Piskunov (FKPP) equation provides a
paradigmatic example of front propagation and asymptotic-state selection.
Using a unified renormalization-group (RG) framework, we show that its
traveling-wave solutions form a continuous family of RG fixed points and
that all fronts with $v\geq 2$ are locally RG-stable. The resulting RG
stability criterion reproduces the classical linear-stability analysis of
Canosa, according to which all fronts with $v\geq 2$ are stable.
Nevertheless, the asymptotically selected front is $v=2$, determined by the
leading-edge structure of the initial condition rather than by RG stability.
The FKPP equation thus provides an explicit example in which RG fixed-point
stability does not determine asymptotic-state selection, demonstrating that
stability and selection constitute distinct aspects of RG behavior.

\end{abstract}

\maketitle

\section{Introduction}

Traveling waves are among the most important asymptotic structures in
nonlinear dynamics and arise in a wide variety of systems, including
reaction--diffusion equations, biological invasions, combustion waves, and
chemical reactions \cite{murray2002mathematical1, murray2002mathematical2,
canosa1973, Barenblatt}. Among the many equations exhibiting traveling-wave
solutions, the Fisher--Kolmogorov--Petrovsky--Piskunov (FKPP) equation
occupies a distinguished position because it provides one of the simplest
and best-studied examples of front propagation and front selection \cite%
{fisher1937wave, kpp1937_collected, canosa1973, murray2002mathematical1,
murray2002mathematical2, Barenblatt}. Originally introduced to describe the
spatial spread of advantageous genes in a population \cite{fisher1937wave,
kpp1937_collected}, the FKPP equation has since been widely used to model
biological invasions, epidemic spreading, combustion fronts, and
autocatalytic chemical reactions \cite{murray2002mathematical1,
murray2002mathematical2}.

Classical analyses established the existence of traveling-wave solutions of
the FKPP equation and clarified their propagation and minimal-speed
properties \cite{kpp1937_collected,
aronson1975nonlinear,aronson1978multidimensional}. Later, Canosa
investigated the linear stability of the resulting traveling-wave family 
\cite{canosa1973}. A distinct question concerns which member of this family
is dynamically selected. The asymptotically selected front was found to
depend sensitively on the structure of the leading edge of the initial
condition. Kolmogorov, Petrovsky, and Piskunov showed that compactly
supported initial conditions asymptotically select the minimal velocity $v=2$
\cite{kpp1937_collected}. More generally, McKean and Larson demonstrated
that the selected velocity is determined by the exponential decay rate of
the leading edge \cite{mckean1975,larson1978}. Rigorous analyses further
clarified the asymptotic convergence toward the selected front and the
associated logarithmic corrections to the front position \cite%
{bramson1983convergence}. Subsequent developments interpreted front
selection through marginal-stability arguments \cite{dee1983propagating}.
The distinction between pulled and pushed fronts, together with the central
role of leading-edge dynamics in front propagation and front selection, was
later established in a unified framework \cite{ebert2000front} and
comprehensively reviewed by van Saarloos \cite{van2003front}. These studies
established front selection as one of the central problems in the theory of
nonlinear waves and highlighted the distinction between selection and
stability.

A related question concerns the robustness of the selected front under
perturbations of the governing equation. This issue was addressed through
the concept of structural stability \cite{paquette1994structural}, according
to which physically observable fronts are identified with fronts that remain
stable against physically small perturbations of the governing equation \cite%
{chen1994rgfronts,chen1994selection}. Structural stability was further
proposed as an alternative front-selection principle applicable to both
pulled and pushed fronts \cite{paquette1994structural,chen1994rgfronts,
chen1994selection}.

Renormalization-group (RG) methods provide a powerful framework for
understanding asymptotic behavior in nonlinear systems. Motivated by
Barenblatt's theory of self-similarity, Goldenfeld and co-workers developed
field-theoretic RG approaches to nonlinear partial differential equations
and demonstrated that self-similar solutions of the second kind can be
understood in terms of anomalous dimensions \cite%
{goldenfeld1989intermediate, goldenfeld1990anomalous,
chen1994renormalization, chen1995numerical,Goldenfeld}. Independently,
Bricmont, Kupiainen, and Lin introduced a Wilson-type RG formulation for
nonlinear parabolic equations \cite{bricmont1994renormalization}, while Giga
and Kohn and others developed a dynamical-systems description of
asymptotically self-similar solutions \cite{giga1985asymptotically,
eggers2015singularities}. More recently, these three approaches were unified
within a common RG framework for nonlinear PDEs \cite{Okumura2025RG,
okumura2026combined, Okumura2026oil, Okumura2026nonlinear}. This framework
identifies asymptotic solutions with RG fixed points and constructs
universality classes through the progressive elimination of irrelevant
structures under repeated RG transformations, thereby providing a predictive
criterion for the formation of universality classes \cite%
{Okumura2026nonlinear}.

More recently, applications of the unified framework have identified two
further possibilities: memory-retaining fixed points \cite{Okumura2026memory}
and traveling-wave fixed points without universality classes \cite%
{Okumura2026traveling}. These results demonstrate that the existence of RG
fixed points does not necessarily imply either complete information loss or
the formation of universality classes. Having established the existence of
traveling-wave fixed points, a natural next question concerns their
stability and dynamical significance. Within RG theory, fixed-point
stability is often closely associated with asymptotic behavior. It is
therefore natural to ask whether asymptotic-state selection can also be
understood solely from RG fixed-point stability. The FKPP equation provides
an ideal testing ground because both front selection and traveling-wave
stability have been extensively studied. In particular, does RG fixed-point
stability determine which asymptotic state is selected dynamically?

The purpose of the present work is to address this question for the FKPP
equation. We show that the family of classical traveling-wave solutions
emerges naturally as a continuous family of RG fixed points. Furthermore,
the linear stability analysis of the corresponding RG flow reproduces the
classical stability analysis of Canosa \cite{canosa1973} and predicts that
all fronts with $v\geq 2$ are locally stable. RG stability therefore does
not determine which front is asymptotically selected. Instead, the
asymptotically selected front is $v=2$, determined by the leading-edge
structure of the initial condition. The FKPP equation therefore provides an
explicit example in which RG fixed-point stability and front selection are
distinct concepts, thereby clarifying both the scope and the limitations of
RG descriptions of asymptotic dynamics.\newline

\section{Unified RG framework}

The present analysis is based on the previously developed RG framework for
PDE of the form \cite{Okumura2025RG, okumura2026combined, Okumura2026oil,
Okumura2026nonlinear}%
\begin{equation}
\partial _{T}\mathbf{H}(T;\mathbf{X})=\mathbf{F}(\mathbf{H},D_{1}\mathbf{H}%
,D_{2}\mathbf{H},\cdots )+\mathbf{G}(\mathbf{H},D_{1}\mathbf{H},D_{2}\mathbf{%
H},\cdots )  \label{eq13}
\end{equation}%
In the RG framework, a scale transformation for the scale factor $L$ and the
exponent $B$, which are both positive, is defined as

\begin{eqnarray}
T^{\prime } &=&T/L^{B},\text{ }\mathbf{X}^{\prime }=\mathbf{X}/L  \notag \\
H_{i}^{\prime }(T^{\prime };\mathbf{X}^{\prime }) &=&L^{A_{i}}H_{i}(T,%
\mathbf{X})\text{ }\equiv H_{i}^{L}(T^{\prime },\mathbf{X}^{\prime })
\label{eq15} \\
&\Leftrightarrow &\text{ }H_{i}^{L}(T,\mathbf{X})=L^{A_{i}}H_{i}(L^{B}T,L%
\mathbf{X}).  \label{eq15C}
\end{eqnarray}%
\newline

The scale factor $L$ satisfies $L>1$ for \textit{long-time asymptotics} in
the present case where \textit{large scale physics} is important (Case II,
in the previous study \cite{Okumura2026nonlinear}). The terms in $\mathbf{F}$
and $\mathbf{G}$ are regarded as \textit{scale-invariant}, \textit{relevant}%
, and \textit{irrelevant} depending on whether the exponent $M$ defined in
Eq \ref{M1} is zero, positive, and negative, respectively, for Case II.
Further details of the unified RG framework are given in Appendix \ref{A-U}.

In the following, the terms in $\mathbf{F}$ are scale invariant, whereas the
terms collected in $\mathbf{G}$ are non-scale-invariant and are assumed here
to be irrelevant. Accordingly, the terms in $\mathbf{F}$ are retained,
whereas those in $\mathbf{G}$ are eliminated as the RG transformation
defined in Eq \ref{eq17} is repeatedly applied. Equation \ref{eq13} with $%
\mathbf{G=0}$ defines a universality class, while equations with nonzero
irrelevant $\mathbf{G}$ belong to the same universality class. Thus,
universality arises when the non-scale-invariant contribution $G$ is
eliminated under repeated RG transformations.

\section{\noindent FKPP equation}

We consider the long time asymptotics of FKPP equation:%
\begin{equation}
\partial _{t}V(t,x)=V(1-V)+\partial _{x}^{2}V  \label{eq01}
\end{equation}%
In the spatially homogeneous case, the steady states are $V=0$ and $V=1$,
which are respectively unstable and stable. This suggests that we should
look for travelling waves satisfying the following conditions: 
\begin{eqnarray}
0 &\leq &V\leq 1  \label{eq1A} \\
V(0,x) &\rightarrow &\left\{ 
\begin{array}{ccc}
0 &  & x\rightarrow \infty \\ 
1 &  & x\rightarrow -\infty%
\end{array}%
\right.  \label{eq1B}
\end{eqnarray}%
We seek a traveling-wave solution of the form, i.e., the following
traveling-wave ansatz: 
\begin{equation}
V(t,x)=\widehat{V}(\zeta =x-vt+k)  \label{eq01c}
\end{equation}%
with the front moving with a velocity $v$.

As shown previously for other types of traveling waves \cite%
{Okumura2026traveling}, the logarithmic transformation maps the FKPP front
onto a self-similar RG fixed point with vanishing field dimension $A=0$. To
briefly see this, we introduce variables: $t=\log T-\log T_{0}$; $x=\log X$.
We then obtain $x-vt+k=\log (\frac{X}{KT^{v}})$ with $T_{0}^{-v}=K$. For the
new variables $T$ and $X$ with $V(t,x)=H(T,X)$, \textit{the traveling-wave
solution can be regarded as a self-similar solution}, $H(T,X)=\widehat{H}(%
\frac{X}{KT^{v}})$, a special case of Eq \ref{e20} with $\bm{\xi }%
\rightarrow X/T^{1/B}$ and $A=0$. The vanishing field dimension $A=0$ is
further justified by the scaling analysis given below. On the other hand,
with $T$ and $X$, FKPP equation changes into the following from%
\begin{equation}
T\partial _{T}H=F[H]  \label{eq03}
\end{equation}%
with

\begin{equation}
F[H]=H(1-H)+(X\partial _{X}H+X^{2}\partial _{X}^{2}H).  \label{eq04}
\end{equation}

The replacements, $T\rightarrow L^{B}T$ and $X\rightarrow LX$, with $%
H_{L}(T,X)=L^{A}H(L^{B}T,LX)$, result in the following transformations: $%
T\partial _{T}H\rightarrow L^{-A}H_{L}$, $H^{n}\rightarrow L^{-nA}H_{L}$,
and $X^{n}\partial _{X}^{n}H\rightarrow L^{-A}X^{n}\partial _{X}^{n}H_{L}$
with $n=1,2,\ldots $. Within the unified RG framework, asymptotic solutions
correspond to RG fixed points and the existence of a fixed point requires Eq %
\ref{eq04} to be scale invariant. This requirement is fulfilled in the
present case only when $A=0$. This in turn leads to the absence of
universality classes, as discussed previously for traveling-wave systems 
\cite{Okumura2026traveling}.

By applying the RG transformation for $L>1$ repeatedly to investigate the RG
flow as explained in Appendix \ref{A-U}, we find from Eq \ref{e20} that the
unified RG framework gives the RG-fixed-point self-similar solution as%
\begin{equation}
H^{\ast }(T,X)=h^{\ast }(\xi )\text{ with }\xi =\frac{X}{KT^{1/B}}.
\label{eq05}
\end{equation}%
Introducing the logarithmic time $\tau =B\log L$, and taking the
continuous-flow limit described in Appendix \ref{A-U}, we obtain $B\frac{%
dH_{L}(T,X)}{d\tau }$ $=AH_{L}$ $+X\frac{\partial H_{L}}{\partial X}$ $%
+BF[H_{L}]$, in which we set $T=1$ and $H_{L}(1,X)=R_{L}f(X)\equiv \widehat{h%
}(\tau ,X)$, to get the RG flow equation: 
\begin{equation}
B\frac{d\widehat{h}(\tau ,X)}{d\tau }=X\frac{\partial \widehat{h}}{\partial X%
}+BF[\widehat{h}],  \label{eq07}
\end{equation}%
which corresponds to Eq \ref{eq21}.

The stationary solution $\widehat{h}(\tau ,X)=f(X)$ to the flow equation
when $\frac{d\widehat{h}(\tau ,X)}{d\tau }=0$ satisfies the nonlinear
second-order ordinary differential equation: 
\begin{equation}
vXf^{\prime }+f(1-f)+(Xf^{\prime }+X^{2}f^{\prime \prime })=0,  \label{eq08}
\end{equation}%
with $1/B=v$.

Equation \ref{eq08} is the RG fixed-point equation obtained as the
stationary solution to the RG flow equation, which leads to the equation, $v%
\widetilde{f}^{\prime }+\widetilde{f}(1-\widetilde{f})+\widetilde{f}^{\prime
\prime }=0$ through the introduction of the function $\widetilde{f}(x)=f(X)$
with $x=\log X$, recovering the results that appear in the standard
literature (e.g., \cite{canosa1973, murray2002mathematical1}).

No analytical solutions to this equation, equivalent to Eq \ref{eq08}, for
general $v$ have been found, although, based on a perturbation\ in terms of $%
\varepsilon =1/v^{2}$, Canosa obtained an asymptotic solution (see Appendix %
\ref{AC}).

In the present study, we are rather interested in the stability of a
full-order solution of Eq \ref{eq08}, $f=f_{v}$, by considering a small
perturbation:%
\begin{equation}
\widehat{h}(\tau ,X)=f_{v}(X)+\delta (X)e^{\omega \tau }
\end{equation}%
From Eq \ref{eq07}, we obtain%
\begin{equation}
\omega \delta =vX\delta ^{\prime }+(1-2f_{v})\delta +(X\delta ^{\prime
}+X^{2}\delta ^{\prime \prime })  \label{c3}
\end{equation}%
up to the first order in $\delta $.

Equation \ref{c3} is obtained as the equation governing the linear stability
of RG fixed points, and as expected, recovers the standard stability
analysis \cite{canosa1973, murray2002mathematical1} as follows. Rewriting
this with the old variable $x=\log X$, we obtain an equation for $%
d(x)=\delta (X)$ with $\widehat{f}_{v}(x)=f_{v}(X)$:%
\begin{equation}
d^{\prime \prime }+vd^{\prime }+(1-\omega -2\widehat{f}_{v})d=0.
\label{eq14}
\end{equation}%
Introducing the variable $g(x)=e^{-cvx}d(x)$, we further obtain $d^{\prime
}=e^{cvx}(cvg+g^{\prime })$ and $d^{\prime \prime }=e^{cvx}(cv(cvg+g^{\prime
})+(cvg^{\prime }+g^{\prime \prime }))$, and thus $(c^{2}v^{2}g+2cvg^{\prime
}+g^{\prime \prime })$ $+v(cvg+g^{\prime })$ $+$ $(1-\omega -2f_{v})$ $g$ $%
=0 $. Setting $c=-\frac{1}{2}$ for the coefficient of $g^{\prime }$ to
vanish, we then obtain a differential equation of Sturm-Liouville type:

\begin{equation}
g^{\prime \prime }+(\lambda +q)g=0.  \label{eq16}
\end{equation}%
with $-\lambda =\omega $ and $-q(x)=2\widehat{f}_{v}+(\frac{v}{2})^{2}-1$.
If we further consider an initial-value problem with a compact condition:%
\begin{equation}
\left\{ 
\begin{array}{ccc}
V(0,x)=1 &  & x\leq -L \\ 
0<V(0,x)<1 &  &  \\ 
V(0,x)=0 &  & x\geq L%
\end{array}%
\right.  \label{eq18}
\end{equation}%
Eq \ref{eq16} should be solved with the boundary condition $g(x=\pm L)=0$,
which defines an eigenvalue problem for $\lambda $. Then, it is known that $%
\lambda $ is real and positive if $-q(x)$ $=$ $2\widehat{f}_{v}$ $+(\frac{v}{%
2})^{2}$ $-1$ $\geq $ $0$ (see Appendix \ref{A-SL}), which is guaranteed
when $v\geq 2$, since $\widehat{f}_{v}(x)\geq 0$ based on Eq \ref{eq1A}.
Within this local stability problem, $\widehat{f}_{v}(x)$ is stable for $%
v\geq 2$.

The stability analysis presented above reproduces that of Canosa in 1973 
\cite{canosa1973}. However, the key equation \ref{eq14} is obtained here
from the linear stability analysis of the RG fixed point, which is
determined as a stationary self-similar solution to the RG flow equation,
rather than from a direct stability analysis of the traveling-wave solution
in its conventional (non-self-similar) representation. The agreement with
Canosa's classical linear-stability criterion is therefore not an accidental
coincidence. Rather, it shows that, in the present setting, the local
stability of the RG fixed point provides an RG reinterpretation of the
classical linear stability of the corresponding traveling wave.

It is known that the translating velocity $v$ is critically dependent on the
initial conditions at infinity. For example, Mollison showed \cite%
{Mollison1977} that, for the leading-edge solution of the form $V(t,x)\sim
e^{-\alpha (x-vt)}$, the velocity $v$ satisfies the dispersion relation: $%
v=\alpha +\frac{1}{\alpha }$ (see Appendix \ref{A-2}), which has a minimum $%
v_{0}=2$. This relation establishes a direct connection between the
exponential decay rate $\alpha $ of the leading edge and the propagation
velocity. McKean \cite{mckean1975} and Larson \cite{larson1978} subsequently
showed that the asymptotically selected velocity is determined by this
leading-edge decay rate, yielding $v=\alpha +\frac{1}{\alpha }$ for $0\leq
\alpha \leq 1$ and $v=2$ for $\alpha \geq 1$. In particular, sufficiently
steep leading edges ($\alpha \geq 1$) select the minimal velocity $v=2$. In
addition, Kolmogoroff, Petrovsky, and Piskunov proved that $v=2$ if $V(0,x)$
satisfies Eq \ref{eq18}. Rigorous analyses further established convergence
toward the selected front and clarified logarithmic corrections to the front
position \cite{bramson1983convergence}. These results were later interpreted
through marginal-stability arguments, according to which front selection is
governed by the dynamics of the leading edge and the selected velocity
corresponds to the minimum of the dispersion relation $v=\alpha +\frac{1}{%
\alpha }$ \cite{dee1983propagating}. The distinction between pulled and
pushed fronts, together with the central role of leading-edge dynamics in
front propagation and front selection, was subsequently established within a
unified framework \cite{ebert2000front} and comprehensively reviewed by van
Saarloos \cite{van2003front}.

The RG analysis determines the existence and local stability of the
traveling-wave fixed-point family, but not the asymptotically selected
front. The latter is controlled by the leading-edge structure of the initial
condition.\newline

\section{Discussion}

Within the unified RG framework, traveling-wave solutions of the FKPP
equation form a continuous family of RG fixed points. The FKPP equation
therefore provides a useful setting for comparing RG fixed-point stability
with asymptotic-state selection.

\textit{The present RG stability analysis reproduces the classical
linear-stability result of Canosa. In the present setting, this agreement is
expected from the interpretation of traveling-wave solutions as RG fixed
points. The RG fixed-point stability analysis may therefore be viewed as a
renormalization-group reinterpretation of the classical linear-stability
criterion.}

\textit{At the same time, the present analysis shows that such stability
alone does not determine front selection: all fronts with }$v\geq 2$\textit{%
\ are locally RG stable, whereas only one front is dynamically selected.
Front selection should therefore be regarded as a problem complementary to,
but distinct from, RG fixed-point stability. }

A further question concerns the robustness of the selected front under
perturbations of the governing equation. This issue was addressed through
the concept of structural stability \cite%
{paquette1994structural,chen1994rgfronts, chen1994selection}. Although
structural stability also involves perturbations, it addresses a different
question. Structural stability concerns the effect of physically small
perturbations of the governing equation itself, whereas the present work
concerns the local stability of RG fixed points within a fixed equation. In
particular, the $v=2$ front is structurally stable, whereas the $v>2$ fronts
are not \cite{paquette1994structural, chen1994selection}. Thus, structural
stability distinguishes the selected $v=2$ front from the locally stable $%
v>2 $ fronts. Moreover, structural stability should be distinguished from RG
analyses of universality classes, which examine the asymptotic fate of
perturbations added to the governing equation under repeated RG
transformations rather than their direct effect on the observable front.

A separate distinction concerns universality-class analysis. In particular,
whereas Ref \cite{Okumura2026traveling} showed that traveling-wave fixed
points may exist without generating universality classes, the present work
examines the stability and asymptotic-state selection properties of such
fixed-point families. In systems possessing universality classes, repeated
RG transformations eliminate irrelevant structures and drive a broad range
of initial conditions toward the same asymptotic state. By contrast,
traveling-wave fixed points do not generate universality classes within the
unified RG framework. The FKPP equation suggests that, in such situations,
the asymptotic state may not be determined solely by RG fixed-point
stability, leaving room for additional selection mechanisms.

More broadly, together with the previously identified cases of
memory-retaining fixed points \cite{Okumura2026memory} and traveling-wave
fixed points without universality classes \cite{Okumura2026traveling}, the
present result suggests that RG fixed points, universality classes,
fixed-point stability, and asymptotic-state selection constitute distinct
aspects of asymptotic RG behavior. The FKPP equation provides an explicit
example in which a continuous family of stable RG fixed points exists, yet
the asymptotic state is determined by a separate selection mechanism.

\section{\noindent \noindent Conclusion\label{Conclusion}}

We have shown that the RG stability analysis of the FKPP equation reproduces
the classical linear-stability result of Canosa while leaving front
selection undetermined. The stability discussed here refers to local
stability of the corresponding RG fixed points. The unified RG framework
determines the existence and local stability of traveling-wave fixed points,
whereas the selected asymptotic front is governed by leading-edge dynamics.
The FKPP equation therefore identifies a fundamental limitation of RG
descriptions based solely on fixed points and their local stability.

\textit{The author is grateful to Professor Nigel Goldenfeld (UCSD) for
helpful comments, particularly regarding the distinction between RG
fixed-point stability and structural stability. This work was supported by
JSPS\ KAKENHI Grant Number JP24K00596. }

\appendix
\clearpage

\section{Supplementary notes\label{A2}}

\subsection{Note on the unified RG framework\label{A-U}}

We provide here additional information on the unified RG framework \cite%
{Okumura2026nonlinear}. The term in $\mathbf{F}$ and $\mathbf{G}$ in Eq \ref%
{eq13} of the form,%
\begin{equation}
H_{1}^{N_{1}}H_{2}^{N_{2}}\cdots (\partial _{X_{1}}H_{1})^{N_{11}}(\partial
_{X_{1}}H_{2})^{N_{12}}\cdots =\dprod\limits_{i,j,k,l,n,\cdots
}H_{i}^{N_{i}}(\partial _{X_{j}}H_{k})^{N_{jk}}(\partial _{X_{l}}\partial
_{X_{m}}H_{n})^{N_{lmn}}\cdots ,  \label{t1}
\end{equation}%
is changed into the form, $L^{M}\tprod\limits_{i,j,k,l,n,\cdots }\left[
H_{i}^{L}\right] ^{N_{1}}(\partial _{X_{j}}H_{k}^{L})^{N_{jk}}(\partial
_{X_{l}}\partial _{X_{m}}H_{n}^{L})^{N_{lmn}}\cdots $, under the scale
transformation given in Eq \ref{eq15}. Here, we introduced the \textit{%
scaling factor} $L^{M}$ characterized by the \textit{scaling exponent} $M$,
which defines relevance of the term:%
\begin{equation}
M=A_{I}+B-[\sum_{i}N_{i}A_{i}+\sum_{j,k}N_{jk}(A_{k}+1)+%
\sum_{l,m,n}N_{lmn}(A_{n}+2)+\cdots ],  \label{M1}
\end{equation}

In this study, we are interested in \textit{intermediate asymptotics} (Case
II in the previous article \cite{Okumura2026nonlinear}) with the boundary
condition at $T=1$: $\mathbf{H}(1,\mathbf{X})=\mathbf{h}(\mathbf{X})$. The
RG transformation is defined for the $i$-th component $h_{i}(\mathbf{X})$ of 
$\mathbf{h}(\mathbf{X})$ as%
\begin{equation}
\emph{R}_{L,\mathbf{G}}h_{i}(\mathbf{X})\equiv L^{A_{i}}H(L^{B};L\mathbf{X}%
)=H_{i}^{L}(1;\mathbf{X})\text{,}  \label{eq17}
\end{equation}%
where the second equality is due to Eq \ref{eq15C}.

If $\mathbf{F}$ is scale-invariant and $\mathbf{G}$ is irrelevant, we can
expect that iteration of RG makes $\mathbf{h}(\mathbf{X})$ and Eq \ref{eq13}
flow into their fixed points: $\mathbf{h}^{\ast }(\mathbf{X})$ and 'Eq \ref%
{eq13} with zero $\mathbf{G}$,' where the fixed point is defined by the
following equation: 
\begin{equation}
\emph{R}_{L,\mathbf{G}^{\ast }}\mathbf{h}^{\ast }(\mathbf{X})=\mathbf{h}%
^{\ast }(\mathbf{X})\Leftrightarrow L^{A_{i}}H_{i}^{\ast }(L^{B};L\mathbf{X}%
)=h_{i}^{\ast }(\mathbf{X}).
\end{equation}%
If such a point exists, setting $T=L^{B}$ in this equation results in $%
T^{A_{i}/B}H_{i}(T;T^{1/B}\mathbf{X})=\mathbf{h}^{\ast }(\mathbf{X})$, from
which we obtain a self-similar solution: 
\begin{equation}
H_{i}^{\ast }(T;\mathbf{X})=T^{-A_{i}/B}h_{i}^{\ast }(\bm{\xi })\text{ with }%
\bm{\xi }=\mathbf{X}/T^{1/B}  \label{e20}
\end{equation}

To seek fixed points and their stability, it is convenient to apply the RG
transformation successively $n$ times with scale factor $L$. When the
governing equation is scale invariant, the successive transformations are
equivalent to a single RG transformation with scale factor $L^{n}$. By
introducing the logarithmic time $\tau =B\log L$ for Case II, we have $\tau
_{n}=B\log L^{n}=n\tau $. Thus, the multiplicative growth of the RG scale is
represented by an additive parameter, allowing repeated RG transformations
to be viewed as a continuous flow. Taking the infinitesimal-step limit $%
L\rightarrow 1$ yields the RG flow equation for $\widehat{\mathbf{h}}(\tau ;%
\mathbf{X})\equiv \mathbf{H}_{L}(1;\mathbf{X})=\emph{R}_{L}\mathbf{f}(%
\mathbf{X})$ :

\begin{equation}
B\frac{d\widehat{h_{i}}(\tau ;\mathbf{X})}{d\tau }=A_{i}\widehat{h_{i}}(\tau
;\mathbf{X})+BF_{i}(\widehat{\mathbf{h}},D_{1}\widehat{\mathbf{h}},D_{2}%
\widehat{\mathbf{h}},\cdots )+\sum_{i}X_{i}\frac{\partial \widehat{h_{i}}%
(\tau ;\mathbf{X})}{\partial X_{i}}.  \label{eq21}
\end{equation}

When a non-scale-invariant contribution $G$ is irrelevant under the RG flow,
repeated RG transformations progressively suppress its influence, and the
dynamics approaches the scale-invariant equation governed by $F$.
Universality classes are thus generated through the repeated elimination of
irrelevant structures.

By setting $\frac{d\widehat{h_{i}}(\tau ;\mathbf{X})}{d\tau }=0$ in Eq \ref%
{eq21}, we obtain the fixed-point equation, 
\begin{equation}
0=A_{i}h_{i}^{\ast }(\mathbf{X})+BF_{i}(\mathbf{h}^{\ast },D_{1}\mathbf{h}%
^{\ast },D_{2}\mathbf{h}^{\ast },\cdots )+\sum_{i}X_{i}\frac{\partial
h_{i}^{\ast }(\mathbf{X})}{\partial X_{i}},  \label{eq21a}
\end{equation}%
for the stationary solution $\mathbf{h}^{\ast }(\mathbf{X})$. Its stability
can be examined by substituting 
\begin{equation}
\widehat{\mathbf{h}}(\tau ;\mathbf{X})=\mathbf{h}^{\ast }(\mathbf{X})+%
\bm{\delta}(\mathbf{X})e^{\omega \tau }
\end{equation}%
into Eq \ref{eq21}, linearizing the equation in terms of $\bm{\delta}(%
\mathbf{X})$, and examining the sign of $\omega $. Since $\tau $ goes to
positive infinity for both Cases of I and II with repeated application of RG
transformation, a negative $\omega $ corresponds to a \textit{stable}
(decaying) mode, whereas a positive $\omega $ indicates an \textit{unstable}
(growing) mode. The case $\omega =0$ corresponds to a \textit{marginal}
mode, which may be absorbed into the fixed point solution depending on the
structure of the problem.

\subsection{Note on the perturbative analytical solution\label{AC}}

The leading-order asymptotic solution is given as: $f^{(0)}\equiv f(X)=\frac{%
1}{1+X^{1/v}}$. In the present context, we can obtain this as follows. By
further introducing the function $\widehat{f}(\xi )=\widetilde{f}(x)$ with $%
\xi =x/v=\varepsilon ^{1/2}x$, we obtain the equation $\widehat{f}^{\prime }+%
\widehat{f}(1-\widehat{f})+\varepsilon \widehat{f}^{\prime \prime }=0$. In
the small $\varepsilon $ limit, the latter differential equation has the
solution $\widehat{f}(\xi )=$ $\frac{1}{1+e^{\xi }}$ with the condition $%
\widehat{f}(0)=1/2$. This is the leading order solution $f^{(0)}$ given
above.

\subsection{Note on the eigenvalue of Sturm-Liouville problem\label{A-SL}}

From Eq \ref{eq16}, we can prove $-\int_{-L}^{L}dxgg^{\prime \prime
}-\int_{-L}^{L}dxqg^{2}=\lambda \int_{-L}^{L}dxg^{2}$, which shows that $%
\lambda $ is positive if the left-hand side is positive. This follows from $%
q(x)\leq 0$ and the identity $\int_{-L}^{L}dx(g^{\prime })^{2}=gg^{\prime
}|_{-L}^{L}-\int_{-L}^{L}dxgg^{\prime \prime }$ with $g(x=\pm L)=0$.

\subsection{Note on the leading-edge theory\label{A-2}}

We consider the leading edge of the evolving wave where the linearized
equation$\ \partial _{t}V(t,x)=V+\partial _{x}^{2}V$ is valid since $%
V^{2}\ll V$ at the edge. Substituting this equation $V(t,x)\sim e^{-\alpha
(x-vt)}$, consistent with $V(0,x)\sim e^{-\alpha x}$, we obtain $v\alpha
=1+\alpha ^{2}$, i.e., $v=\alpha +\frac{1}{\alpha }$, which has a minimum at 
$v_{0}=2$ at $\alpha =1$.

\end{document}